\documentclass[conference]{IEEEtran}
\IEEEoverridecommandlockouts
\usepackage{cite}
\usepackage{amsmath,amssymb,amsfonts}
\usepackage{algorithmic}
\usepackage{graphicx}
\usepackage{textcomp}
\usepackage{xcolor}
\usepackage{tcolorbox}
\usepackage{booktabs} 
\def\BibTeX{{\rm B\kern-.05em{\sc i\kern-.025em b}\kern-.08em
    T\kern-.1667em\lower.7ex\hbox{E}\kern-.125emX}}

\newcommand{\rev}[1]{\textcolor{black}{#1}}

\begin{document}

\title{Revisiting Feedback-Driven LLM Code Repair: A Replication and Exploratory Java Extension}

\author{
\IEEEauthorblockN{Louis Lalonde, Wassim Keddache,
Thomas Perron Touchette, Leuson Da Silva, and Foutse Khomh}
\IEEEauthorblockA{
Polytechnique Montr\'eal, Montr\'eal, Canada\\
\{louis.lalonde, wassim.keddache, thomas.perron-touchette, leuson-mario-pedro.da-silva, foutse.khomh\}@polymtl.ca}
}

\maketitle

\begin{abstract}
Since the advent of Large Language Models (LLMs), practitioners have increasingly leveraged them to support their software engineering tasks, including automated code repair, showing promising results. 
Yet, concerns regarding reproducibility and generalizability remain largely unexplored. 
To further evaluate these concerns and associated impacts, we partially reproduce and conduct an exploratory Java extension of the FeedbackEval benchmark \cite{dai2025feedbackeval}, which evaluates how LLMs leverage different feedback types for Python code repair. 
First, we partially replicate the original study on 394 repair tasks using GPT-4o and Claude 3.5 Sonnet, reproducing and observing the main qualitative trends reported in the original work.
Second, we conduct an exploratory Java extension by constructing 100 erroneous repair instances from 50 Java tasks and evaluating feedback effectiveness.
Our results show that previous conclusions from Python may be sensitive to benchmark construction, feedback representation, and tooling ecosystem, motivating more controlled multilingual benchmarks. 
Specifically, while \textit{test} feedback remains the strongest feedback type in our Python replication, the same ranking is not observed in our Java extension, as \textit{simple} and JUnit-based \textit{test} feedback do not differ significantly. 
We hypothesize that differences in feedback informativeness and tooling ecosystems, such as the verbosity of test frameworks, may partly explain such a difference.
Finally, lighter prompts reduce cost without significant differences in repair effectiveness.
Overall, our findings confirm key trends under a partially controlled replication and highlight the need for more rigorous multilingual evaluation and careful feedback design in LLM-based repair systems.
\end{abstract}


\section{Introduction}
\label{sec-introduction}

During the lifecycle of software projects, bugs and issues are expected to be reported and addressed by developers~\cite{bettenburg2008makes}.
Fixing these issues is often time-consuming and can negatively impact developer productivity~\cite{guo2011not}.
To alleviate this burden, researchers have been extensively studying Automated Program Repair (APR), aiming to automatically generate patches for faulty programs~\cite{le2019automated, zhang2023survey}.
Recent advances in Large Language Models (LLMs), with their ability to understand natural language and generate code, have opened new opportunities for APR.
Prior work has leveraged LLMs for tasks such as bug fixing, vulnerability repair, and integration with static analysis tools~\cite{bouzenia2025repairagent,kulsum2024case,jin2023inferfix}.
However, LLM-based program repair is inherently an iterative process, where models rely on feedback signals to identify and correct errors \cite{peng2025perfcodegen}. 
As a result, understanding how different types of feedback influence repair effectiveness has become a key research question.

Specifically for code repair tasks, previous studies have investigated how different feedback types can improve LLM performance \cite{blyth2025static}.
In this context, the FeedbackEval benchmark \cite{dai2025feedbackeval} demonstrated that, in Python, repair performance varies depending on the feedback type, including unit test execution, static analysis, human feedback, and simple prompts, with \textit{test-based} feedback achieving the highest CoderEval result for GPT-4o (37.8\% Repair@1).
However, whether these findings apply to other programming languages remains an open question~\cite{yang2025survey}.

Programming languages differ fundamentally in their type systems, execution models, and error reporting mechanisms \cite{gu2023self}.
For instance, statically typed languages such as Java enforce type checking at compilation time and produce structured, location-aware diagnostics before execution, while dynamically typed languages, such as Python, detect many errors only at runtime, often resulting in less precise and less actionable feedback~\cite{oh2022pyter,mesbah2019deepdelta}.
Moreover, Java is supported by a mature ecosystem of tools, like \textit{Javac} and \textit{JUnit}, which directly shape how feedback is generated and presented to developers.
These differences may directly influence how LLMs interpret and utilize feedback during repair tasks.
Prior work has also shown that LLMs exhibit varying proficiency across programming languages~\cite{baltaji2025cross,luo2025unlocking}, raising the question of whether feedback-driven repair strategies are efficient across languages with different characteristics.
In addition, practical deployment of LLM-based repair systems requires considering not only effectiveness but also \textit{cost-efficiency} \cite{hidvegi2024cigar}.
While advanced prompting strategies such as \textit{Chain-of-Thought} \cite{wei2022chain} and \textit{Few-shot} \cite{wang2020generalizing} learning have been widely explored, their performance-to-cost trade-offs remain unclear in the context of feedback-driven repair across programming languages~\cite{han2025token,gantayat2025efficiency}.

To address these open questions, we first partly \textbf{replicate} the study by Dai et al.~\cite{dai2025feedbackeval} to assess the reproducibility of its findings in our experimental setup.
We reproduce their evaluation on the CoderEval \cite{yu2024codereval} and HumanEval \cite{chen2021evaluating} benchmarks, considering the initial set of evaluated feedback types (test execution, static analysis, human feedback, and simple prompts) and measuring repair performance using \textit{Repair@1}.
Second, we conduct an \textbf{exploratory extension} of FeedbackEval~\cite{dai2025feedbackeval} to Java.
Specifically, we investigate two research questions: \textbf{(RQ1)} how feedback-effectiveness patterns observed in Python behave under a Java repair setting with different typing, compilation, and testing characteristics; and \textbf{(RQ2)} how cost-efficient different prompting strategies are for feedback-driven Java code repair.
For that, we construct a Java dataset of 100 faulty program instances based on 50 CoderEval tasks by injecting both compiler and logical mutations into correct programs.
We then evaluate LLM-based repair using feedback categories inspired by the original study and evaluate repairs with \texttt{GPT-4o-mini}. 
This design allows us to examine whether the qualitative trends observed in Python are reproduced in our Java extension.
Finally, we analyze the \textit{cost-efficiency} of prompting strategies by evaluating multiple prompting variants, including \textit{Chain-of-Thought} and \textit{Few-shot} prompting, and measuring repair performance relative to API cost.

Our replication reproduces some qualitative FeedbackEval trends, with \textit{test-based feedback} achieving the highest repair
performance in Python. Meanwhile, we observe different results in our exploratory Java extension. For logical tasks, \textit{simple} feedback achieves the highest observed Repair@1, but does not differ significantly from \textit{test} feedback.
For compiler tasks, \textit{compiler} feedback performs best, but does not differ significantly from \textit{simple} feedback.
Finally, lightweight prompting offers more favorable observed cost-effectiveness trade-offs, while \textit{Chain-of-Thought} increases cost without statistically supported gains.

Overall, our findings offer implications for researchers and practitioners.
First, they suggest that conclusions drawn from LLM-based repair studies in a single programming language may be sensitive to its surrounding benchmark, tooling, and feedback-generation setup, highlighting the need for more controlled multilingual evaluation.
Second, our exploratory Java extension results emphasize the role of tooling and feedback quality in shaping LLM behavior, as differences may influence observed repair effectiveness.
Finally, from a \textit{cost-efficiency} perspective, our results indicate that increasing prompt complexity does not necessarily improve performance, underscoring the importance of cost-aware design for scalable LLM-based repair systems.

\section{Replication Study}
\label{sec:replication}

\begin{figure*}[t]
    \centering
     \includegraphics[width=\textwidth, height=0.9\textheight, keepaspectratio]{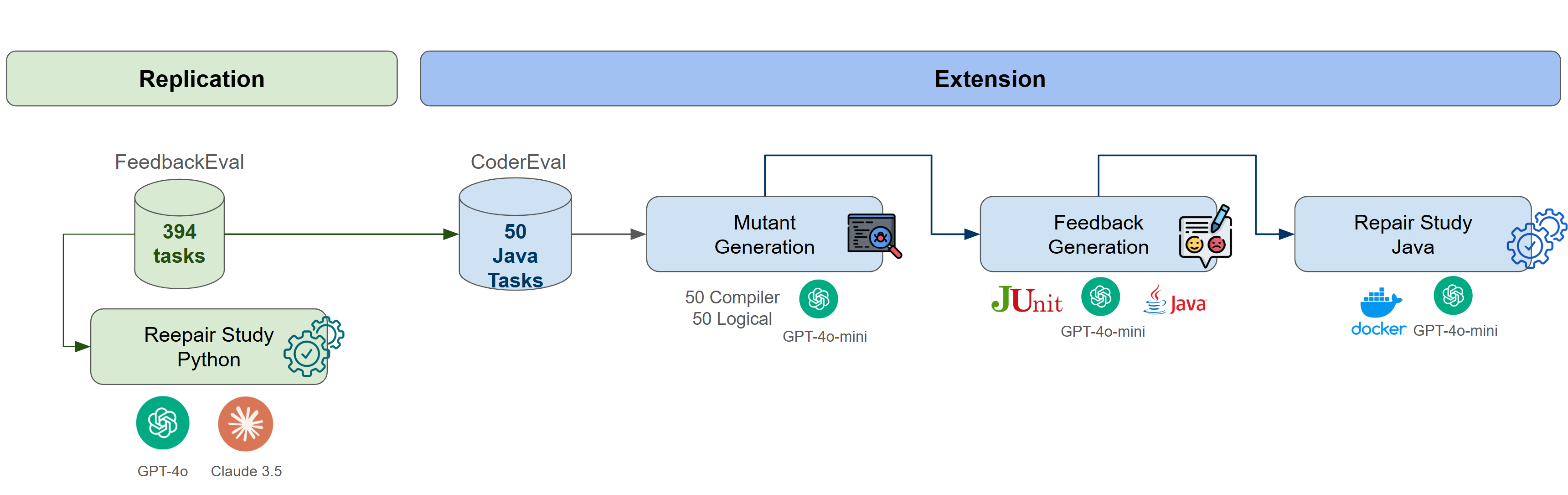}
    \caption{Study Design adopted for our Replication and Extension Study.}
    \label{fig:methodology}
\end{figure*}

In this section, we mainly focus on our replication. 
First, we present the original study conducted by Dai et al.\cite{dai2025feedbackeval}. Then, we present the methodology adopted for our replication, highlighting how we reproduced the experimental setup of the original work and any required adaptations (see Figure \ref{fig:methodology}, left side). 
Second, we present and discuss the results obtained.

\subsection{FeedbackEval: Original Study}
\label{sec-background}

Dai et al. \cite{dai2025feedbackeval} investigate how Large Language Models (LLMs) comprehend and use different types of feedback when performing automated code repair tasks. 
While LLMs have demonstrated strong capabilities in different contexts, like code generation and repair, their ability to effectively leverage different feedback types, ranging from structured test outputs to natural language suggestions, remains poorly understood. 
Aiming to address this gap, the authors introduce FeedbackEval, comprising 394 Python coding tasks drawn from two established benchmarks: 230 tasks from CoderEval \cite{yu2024codereval} and 164 from HumanEval \cite{chen2021evaluating}.\footnote{Our study is based on the first version of the study, available in April, 2025: https://arxiv.org/abs/2504.06939v1}
To convert code-generation problems into repair tasks, FeedbackEval provides 3,736 erroneous instances produced through rule-based mutations, GPT-4o-mini error injection, and incorrect LLM-generated solutions. 
Each task is paired with one erroneous instance and one of four modalities: \textbf{test feedback} from \texttt{pytest}, \textbf{compiler feedback} from \texttt{pylint}, \textbf{human feedback} simulated by GPT-4o-mini, or \textbf{simple feedback}
containing a generic repair request, supporting controlled comparison of feedback-driven repair.

In this context, FeedbackEval is a relevant target for replication because it investigates a timely feedback-driven repair scenario, evaluates multiple feedback strategies, relies on established benchmarks such as CoderEval and HumanEval, and provides reusable artifacts for reproduction.
Overall, these findings highlight the importance of feedback design in guiding LLM-based repair. 

\subsection{Study Design}
\label{sec:study-design-replication}

For our replication, we follow the experimental protocol defined in the original study.
In particular, we reuse the original datasets, buggy code snippets, feedback artifacts, and evaluation scripts made available by the authors. 
We aimed to replicate all evaluated research questions; however, an exact replication was not feasible for RQ3. 
In the original work, 100 tasks were randomly selected from the 230 CoderEval tasks to conduct the iterative feedback experiments. 
However, at the time of our replication, the identifiers of these tasks and the corresponding \texttt{pytest} unit tests used for \textit{test} feedback were not publicly available. 
Reconstructing this subset would require re-randomizing tasks and rewriting tests, which could introduce additional variance and possibly threaten the validity of the comparison. 
Therefore, we exclude RQ3 and replicate the analyses of single-iteration repair across models (RQ1), feedback effectiveness across datasets (RQ2), and prompting techniques (RQ4).

For \textbf{RQ1}, we execute single-iteration code repair experiments. 
Each of the 394 repair tasks is evaluated across the four feedback types (\textit{test, compiler, human, and simple}). 
Repair success is automatically validated by running the unit tests associated with a given task, and results are reported using the Repair@1 metric, representing the proportion of tasks successfully repaired in a single attempt.
For \textbf{RQ2}, we analyze the effectiveness of each feedback modality using the previous execution outputs generated in RQ1. 
Repair success rates are aggregated per feedback type and dataset (HumanEval and CoderEval), allowing us to analyze how different feedback types influence repair performance across models and datasets. 

Finally, for \textbf{RQ4}, we evaluate multiple prompt configurations, including the baseline prompt, \textit{Chain-of-Thought} prompting, \textit{Few-shot} examples, and ablation variations, which remove specific prompt components (e.g., docstring, contextual information, or repair guidelines). 
Each prompt variant is executed on the same set of repair tasks while relying on the same model configuration and feedback type. 
Repair success is also evaluated using the Repair@1 metric, enabling direct comparison with the results reported in the original study.

While the original study evaluates five models, our replication focuses on two representative models: \texttt{GPT-4o} and \texttt{Claude 3.5 Sonnet}. 
This decision was primarily motivated by the substantial computational and financial costs associated with large-scale LLM evaluation, which requires thousands of API calls across multiple configurations.
We selected these models as they represent strong proprietary LLMs from different providers \cite{islam2025gpt, hochmair2024correctness}, enabling us to assess whether the observed trends hold across strong models. 
We also adopt the same decoding parameters whenever possible (temperature = 0.3).

\subsection{Results}
\label{sec:results-replication}
In this section, we present the results of our replication experiments.
For each RQ, we report the findings, comparing them with those reported in the original study. 
Tables \ref{tab:rq1-results}, \ref{tab:rq2-results}, and \ref{tab:rq4-results} summarize the Repair@1 scores across models, feedback types, and experimental settings.

\subsubsection*{RQ1: Single-Iteration Repair Performance}
Overall, our results closely relate to the original results (see Table \ref{tab:rq1-results}).
For \texttt{GPT-4o}, we observe an average Repair@1 score of 55.55\% in our replication, compared to 56.4\% reported in the original paper. 
Regarding performance, it is lowest for \textit{human} feedback and highest for \textit{test} feedback, while \textit{simple} feedback remains competitive across models.
\texttt{Claude 3.5 Sonnet} reports similar behavior with an average Repair@1 score of 60.2\%, which is close to the 60.8\% reported in the original study. 
\textit{Test} feedback again produces the strongest results, achieving 66.6\% Repair@1 in our replication compared to 65.0\% in the original paper.
Overall, these results show that the main qualitative findings are reproducible.
Particularly, we observe that structured feedback, specifically \textit{test} feedback, consistently leads to higher repair success rates than less structured feedback types.

\begin{table}[t]
\centering
\caption{Repair@1 results for single-iteration repair (RQ1).}
\label{tab:rq1-results}
\begin{tabular}{lcc|cc}
\toprule
& \multicolumn{2}{c}{GPT-4o} & \multicolumn{2}{c}{Claude 3.5 Sonnet} \\
Feedback & Replication & Original & Replication & Original \\
\midrule
Human & 50.2 & 50.1 & 56.0 & 55.8 \\
Compiler & 54.1 & 55.6 & 55.9 & 58.9 \\
Simple & 57.5 & 58.2 & 62.5 & 63.5 \\
Test & 60.4 & 61.7 & 66.6 & 65.0 \\
\midrule
Average & 55.55 & 56.4 & 60.2 & 60.8 \\
\bottomrule
\end{tabular}
\end{table}

\subsubsection*{RQ2: Feedback Effectiveness Across Datasets}

For \texttt{GPT-4o}, our results also follow the performance patterns reported in the original study (see Table \ref{tab:rq2-results}).
Repair success is consistently higher on HumanEval than on CoderEval across all feedback types. 
For \textit{CoderEval}, \textit{test} feedback achieves the best performance, while on \textit{HumanEval}, the top results are close, with \textit{simple} feedback slightly exceeding \textit{test} feedback in our replication.
However, we observe some quantitative differences requiring further attention. 
On \textit{HumanEval}, our average Repair@1 score is 76.83\%, slightly lower than the 80.2\% reported in the original study. 
On the other hand, \textit{CoderEval} results are marginally higher (34.27\% compared to 32.6\%).
\texttt{Claude 3.5 Sonnet} exhibits similar dataset-dependent behavior. 
Consistent with the original study, our replication results show that performance is higher on \textit{HumanEval} than on \textit{CoderEval}. Meanwhile, \textit{test} and \textit{simple} feedback also provide the best results. 
Similar to \texttt{GPT-4o}, when the relative ordering of feedback types is preserved, we observe here minor variations for individual scores, specifically for the \textit{compiler} feedback.
Overall, these results confirm that the effectiveness patterns reported in the original study remain largely stable across datasets and models.

\begin{table}[t]
\centering
\caption{Repair@1 results across datasets for GPT-4o (RQ2).}
\label{tab:rq2-results}
\begin{tabular}{lcc|cc}
\toprule
& \multicolumn{2}{c}{HumanEval} & \multicolumn{2}{c}{CoderEval} \\
Feedback & Replication & Paper & Replication & Paper \\
\midrule
Human & 72.0 & 71.6 & 28.5 & 28.5 \\
Compiler & 77.4 & 82.7 & 30.8 & 28.5 \\
Simple & 79.3 & 81.0 & 35.7 & 35.4 \\
Test & 78.7 & 85.6 & 42.1 & 37.8 \\
\midrule
Average & 76.83 & 80.2 & 34.27 & 32.6 \\
\bottomrule
\end{tabular}
\end{table}

\subsubsection*{RQ4: Impact of Prompting Techniques}

Once again, our results show overall consistency with the trends reported in the original study (see Table \ref{tab:rq4-results}). 
We observe that the average Repair@1 score across all prompt configurations
is 50.83\%, closely compared to 50.09\% reported in the original
study.
However, we also observe some differences across individual prompting techniques.
For example, \textit{Chain-of-Thought} reports performance improvements, achieving 53.7\% compared to 52.6\%. 
Conversely, the \textit{baseline} and \textit{Few-shot} configurations perform slightly worse than originally reported.
Some ablation variants report improved results in our replication. 
For example, removing the \textit{docstring} or \textit{persona} components leads to slightly higher
repair success rates. 
In contrast, removing \textit{context information} produces nearly identical results in both studies.
Overall, while quantitative differences are observed across individual prompt configurations, the general conclusion of the original study remains valid: prompt engineering only provides moderate improvements over the \textit{baseline} configuration in single-iteration repair scenarios.

\begin{tcolorbox}[colback=gray!10, colframe=black, boxrule=0.5pt]
\textbf{Summary of Replication.}
Our replication reproduces the key qualitative findings of the original study with minor quantitative differences.
While \textit{test} feedback achieves the highest overall repair success, \textit{simple} feedback remains competitive across models and datasets. 
Finally, the observed quantitative deviations represent a minor concern, falling within expected variability for LLM-based experiments.
\end{tcolorbox}

\begin{table}[t]
\centering
\caption{Repair@1 results across prompting techniques (RQ4).}
\label{tab:rq4-results}
\begin{tabular}{lcc}
\toprule
Prompt Variant & Replication & Paper \\
\midrule
Baseline & 49.5 & 52.6 \\
Chain-of-Thought & 53.7 & 52.6 \\
No Persona & 52.6 & 50.5 \\
Few-shot Example & 48.4 & 50.5 \\
No Docstring & 51.6 & 48.5 \\
No Context & 48.4 & 48.5 \\
No Guideline & 51.6 & 47.4 \\
\midrule
Average & 50.83 & 50.09 \\
\bottomrule
\end{tabular}
\end{table}

\section{Extension Study}
\label{sec-extension}

While our previously reported replication confirms the main qualitative trends reported in the original FeedbackEval study \cite{dai2025feedbackeval}, the target benchmark focuses exclusively on Python repair tasks. 
Such a constraint raises an important question of whether feedback-effectiveness trends observed in Python are reproduced under different programming languages and tooling settings. 
To address this issue, we extend the FeedbackEval benchmark to Java. 
Rather than claiming an isolated language effect, we aim to examine how feedback-driven repair behaves in a Java setting with different compilation, testing, and feedback-generation characteristics.
For that, we construct a controlled set of Java repair tasks and reproduce the experimental setup used in the original benchmark (see Figure \ref{fig:methodology}, right side).

\subsection{RQ1: How do feedback types affect LLM-based repair effectiveness in a Java setting?}

\subsubsection{Motivation}
As previously presented and further validated in our replication, the FeedbackEval benchmark demonstrated that the effectiveness of feedback-driven code repair varies across feedback types in Python.
However, it remains unclear whether the same feedback-effectiveness ranking is reproduced under different programming languages and their tooling settings.
Different languages, such as Java, introduce different characteristics, including explicit type declarations, compilation requirements, and structured compiler diagnostics. 
These properties may affect how LLMs interpret and leverage feedback during automated code repair. 
Therefore, our goal is not to isolate programming language as the only explanatory factor, but to examine whether the qualitative feedback trends observed in Python remain applicable in an exploratory Java repair setting. 

\subsubsection{Approach}
To investigate the effects of the different feedback types analyzed, we extend the original study to Java repair tasks. 
For that, we randomly sample 50 Java tasks from the CoderEval benchmark, which includes tasks derived from real-world projects and has been explored in previous studies. 
Based on the reference correct implementations, we filtered out associated tasks that fail to compile with \texttt{javac} or do not pass the original validation checks, ensuring a reliable baseline for mutation injection.
The validation checks provided in CoderEval are implemented as simple assertion-based scripts rather than full unit test suites executed within a testing framework. 
As a result, failing assertions provide only a binary pass/fail signal and do not expose diagnostic information such as assertion locations, stack traces, or JUnit-style failure reports. 
This limited feedback would provide little guidance for LLM-based repair. 
To address this limitation, we automatically converted the pseudo-unit tests into executable JUnit~4 test suites using \textit{GPT-4o-mini}. 
The conversion was mainly structural: the original assertion scripts were transformed into JUnit test classes, preserving helper methods, setup code, and boolean conditions, while replacing the final pass/fail print logic with explicit \texttt{Assert.assertTrue} checks. 
We manually inspected all 50 translated JUnit tests and found no semantic inconsistencies with the original assertion scripts. 
To further validate test adequacy, all tasks were evaluated by ensuring that (i) correct implementations pass the translated tests and (ii) injected mutations fail them.

While the original study considers three different mutation sources (tool-based, LLM-based, and incorrect code generation), we rely exclusively on LLM-driven mutation injection.
This choice was motivated by recent studies reporting that LLMs can generate realistic and semantically meaningful faulty variants, complementing traditional mutation tools, which often rely on predefined syntactic transformations~\cite{tip2025llmorpheus,wang2026comprehensive}.
On the other hand, such a design choice limits comparability with the original Python benchmark, as the Java mutations may reflect error patterns that are easier for similar LLMs to recognize and repair.
Therefore, we do not interpret the Java extension as a direct cross-language comparison with FeedbackEval. 
Instead, we use it as an exploratory setting to examine how feedback types behave for Java repair tasks. 

For each task, we generate two mutated variants using LLM-driven mutation injection.
The first introduces a \emph{logical mutation}, including incorrect conditions, off-by-one errors, or wrong return values, while the second introduces a \emph{compiler mutation} that prevents successful compilation. 
This process yields 100 erroneous repair instances, 50 per mutation category.
We treat these mutation categories separately because they correspond to two fundamentally different repair scenarios. 
Compiler mutations prevent code execution and require correcting syntactic or typing errors before the program can run. 
In contrast, logical mutations require behavioral corrections that can only be detected through test execution. 
Treating these categories independently allows us to analyze model performance across the two main stages of the repair process for Java. 

Regarding the feedback types, we consider different approaches depending on the mutation category.
For logical mutations, we evaluate three feedback modalities, following the categories defined in the original study~\cite{dai2025feedbackeval}:
\begin{itemize}
\item \textit{test feedback}, derived from executing JUnit test suites and capturing their failure outputs;
\item \textit{human feedback}, simulated using GPT-4o-mini to generate expert-like debugging comments, \rev{following the terminology adopted by FeedbackEval~\cite{dai2025feedbackeval}}; and
\item \textit{simple feedback} (“The code is wrong. Please fix it.”).
\end{itemize}

For compiler mutations, we evaluate \textit{compiler feedback} (derived from \texttt{javac} compilation errors captured during execution), along with \textit{human} and \textit{simple feedback}.
All Java experiments are conducted within a controlled Docker environment to ensure consistent compilation and test execution.

\begin{figure}[t]
    \centering
    \includegraphics[width=\columnwidth]{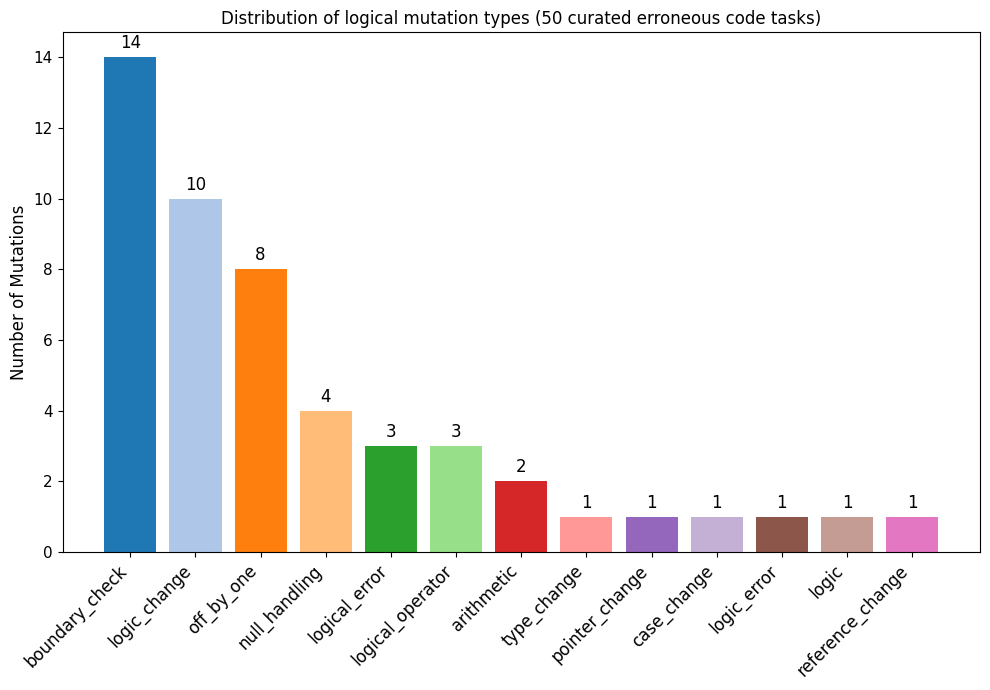}
    \caption{Distribution of logical mutations in the curated 50 Java repair tasks.}
    \label{fig:logical_mutations}
\end{figure}

\textbf{Java Erroneous Code Repair Tasks Dataset Characterization:}
Figure~\ref{fig:logical_mutations} shows the distribution of \textit{logical mutations}, covering error categories such as \textit{boundary condition mistakes}, \textit{off-by-one errors}, \textit{incorrect logic}, and \textit{null handling}.
In contrast, compiler mutations are highly concentrated, with the vast majority corresponding to \textit{syntax} errors (92\%), and a small proportion related to \textit{type} errors (8\%).
These distributions provide important context for interpreting the Repair@1 results, as different mutation types expose distinct feedback signals.
In particular, compiler mutations are dominated by \textit{syntax} errors, which can be effectively addressed using structured compiler diagnostics.
On the other hand, logical mutations are more diverse and require semantic reasoning, making them more sensitive to the quality and expressiveness of feedback, like test outputs vs. simple prompts.
Such a distinction helps explain the variation in feedback effectiveness observed across mutation categories.

\textbf{Metrics and statistical analysis.}
Repair performance is evaluated using \textit{Repair@1}, defined as the percentage of tasks for which the model's first-generated patch produces a program that successfully compiles, when applicable, and passes the associated test suite.
For each Repair@1 estimate, we compute Wilson 95\% confidence intervals.
\rev{Because the same tasks are evaluated under all feedback conditions, we treat the observations as paired.
Within each mutation category, we use Cochran's \(Q\) test to evaluate whether repair outcomes differ overall among the three feedback conditions.
We then conduct pairwise two-sided exact McNemar tests and adjust the three resulting \(p\)-values within each mutation-category family using the Holm procedure.}
We treat differences with an adjusted \(p < 0.05\) as statistically supported.

\subsubsection{Results}

\begin{table}[t]
\centering
\caption{Java Repair@1 results with Wilson 95\% confidence intervals.}
\label{tab:rq1_repair_ci}
\small
\setlength{\tabcolsep}{3pt}
\begin{tabular}{@{}llcc@{}}
\toprule
\textbf{Mutation} & \textbf{Feedback} &
\textbf{Repairs} & \textbf{95\% CI} \\
\midrule
Logical  & Human    & 37/50 (74\%) & [60.45, 84.13] \\
Logical  & Simple   & 43/50 (86\%) & [73.81, 93.05] \\
Logical  & Test     & 41/50 (82\%) & [69.20, 90.23] \\
\midrule
Compiler & Human    & 39/50 (78\%) & [64.76, 87.25] \\
Compiler & Simple   & 46/50 (92\%) & [81.16, 96.85] \\
Compiler & Compiler & 47/50 (94\%) & [83.78, 97.94] \\
\bottomrule
\end{tabular}
\end{table}

Table~\ref{tab:rq1_repair_ci} summarizes the Repair@1 results with Wilson 95\% confidence intervals.
For logical mutations, \textit{simple} feedback achieves the highest
observed Repair@1 (86\%), followed by \textit{test} (82\%) and \textit{human} feedback (74\%).
\rev{Although the conditions differ overall ($Q(2)=8.00$, $p=0.0183$), no pairwise comparison remains significant after Holm correction (all $p_{\mathrm{Holm}}\geq0.0938$).
In particular, the difference between \textit{simple} and \textit{test} feedback is not statistically significant ($p_{\mathrm{Holm}}=0.5000$).
The comparisons between \textit{human} and \textit{simple} feedback ($p_{\mathrm{Holm}}=0.0938$), and between \textit{human} and \textit{test} feedback ($p_{\mathrm{Holm}}=0.4375$), are also not statistically significant.
Therefore, we do not interpret any individual feedback modality as statistically superior for logical mutations.
Instead, the results provide insufficient evidence that the feedback ranking observed in Python persists in our Java setting.}
For compiler mutations, \textit{compiler} feedback achieves the highest observed Repair@1 (94\%), followed by \textit{simple} (92\%) and \textit{human} feedback (78\%).
\rev{The conditions also differ overall ($Q(2)=12.67$, $p=0.0018$).
Both \textit{compiler} (\(p_{\mathrm{Holm}}=0.0234\)) and \textit{simple} feedback (\(p_{\mathrm{Holm}}=0.0313\)) significantly outperform \textit{human} feedback, but do not differ significantly from each other (\(p_{\mathrm{Holm}}=1.0000\)).}

For logical bug repairs, we also observe that some generated patches introduce compilation errors.
This indicates that repair outcomes are not always confined to the original error category: a model, when attempting to fix a semantic issue, may introduce syntactic or type-level problems.
Such an observation suggests that combining multiple feedback signals, such as \textit{test} feedback followed by \textit{compiler} validation, may be useful for more robust repair pipelines.
\subsection{RQ2: How cost-efficient are different prompting techniques when applied to feedback-driven code repair on Java tasks?}

\subsubsection{Motivation}
Advanced prompting strategies such as \textit{Chain-of-Thought} reasoning and \textit{Few-shot} learning have been shown to improve reasoning performance in LLMs \cite{zhang2025enhancing,yu2023towards}. 
However, these techniques typically increase token consumption and API usage costs \cite{nobakhtfar2026chain,li2025cost}. 
In real-world software engineering environments, cost efficiency is often as important as repair effectiveness, especially when automated repair systems must operate at scale. 
The original FeedbackEval benchmark reported only marginal improvements from complex prompting strategies for single-iteration Python repair tasks. 
Given Java’s different typing, compilation, and tooling characteristics, it is unclear whether the same trade-offs hold. 
Understanding the effectiveness–cost balance of prompting strategies is therefore essential for practical deployment of LLM-based repair systems.

\subsubsection{Approach}

\rev{Using the 50 logical-mutation Java tasks constructed for RQ1}, we evaluate different prompting variants similar to the original study~\cite{dai2025feedbackeval}:
(i) \textit{the baseline}, containing the full context (task description, guidelines, docstring, and feedback), 
(ii) \textit{Chain-of-Thought}, encouraging explicit reasoning, and
(iii) \textit{Few-shot}, with a fixed example.
Furthermore, we consider four ablation variants derived from the baseline prompt, each removing a single component:
(i) \textit{No-Docstring} (removal of the function description), 
(ii) \textit{No-Context} (removal of oracle or contextual information such as APIs, variables, and external dependencies), 
(iii) \textit{No-Guidelines} (removal of the repair instruction block), 
and (iv) \textit{No-Persona} (removal of the role-based prompt describing the model as a repair assistant).

For each prompting variant, we execute one repair attempt for each of the same 50 tasks using \texttt{GPT-4o-mini} (temperature=0.3) and record the input and output token usage. 

\textbf{Metrics and statistical analysis.} 
To evaluate cost efficiency, we report:
(i) total API cost,
(ii) Repair@1 with Wilson 95\% confidence intervals,
(iii) cost per successful repair.
Cost per successful repair is the total API cost in CAD divided by the number of successful repairs; lower values indicate greater cost efficiency. 
\rev{Since the same 50 tasks are evaluated under all seven prompting conditions, the repair outcomes are paired at the task level. 
We use Cochran's \(Q\) test to assess whether repair effectiveness differs across the seven conditions. 
We then conduct six planned pairwise comparisons between the baseline and each prompting variant using two-sided exact McNemar tests. 
We adjust the resulting \(p\)-values using the Holm procedure and treat differences with adjusted \(p < 0.05\) as statistically supported.}

\subsubsection{Results}

\begin{table}[t]
\centering
\caption{Cost efficiency of prompting strategies. Costs are reported in CAD, with cost-per-repair computed from unrounded totals.}
\label{tab:prompt_cost_results}
\resizebox{\columnwidth}{!}{%
\begin{tabular}{lcccc}
\toprule
\textbf{Technique}
& \textbf{Repairs}
& \textbf{95\% CI}
& \textbf{Cost}
& \textbf{Cost/Repair} \\
\midrule
Baseline
& 43/50 (86\%) & [73.81, 93.05] & 0.0102 & 0.000236 \\
Chain-of-Thought
& 42/50 (84\%) & [71.49, 91.66] & 0.0241 & 0.000575 \\
Few-shot
& 44/50 (88\%) & [76.20, 94.38] & 0.0122 & 0.000278 \\
No-Context
& 42/50 (84\%) & [71.49, 91.66] & 0.0090 & \textbf{0.000215} \\
No-Docstring
& 40/50 (80\%) & [66.96, 88.76] & 0.0094 & 0.000234 \\
No-Guidelines
& 41/50 (82\%) & [69.20, 90.23] & 0.0090 & 0.000220 \\
No-Persona
& 44/50 (88\%) & [76.20, 94.38] & 0.0099 & 0.000224 \\
\bottomrule
\end{tabular}%
}
\end{table}

Table~\ref{tab:prompt_cost_results} summarizes repair performance and cost efficiency across the prompting strategies. 
\rev{Repair@1 ranges from 80\% to 88\%. 
First, Cochran's \(Q\) test does not identify a statistically significant overall difference among the seven prompting conditions (\(Q(6)=5.04\), \(p=0.5392\)). 
Similarly, none of the six planned comparisons between the baseline and the prompting variants is statistically significant after Holm correction (all \(p_{\mathrm{Holm}}=1.0000\)). 
Thus, we find no statistically supported evidence that any prompting variant changes Repair@1 relative to the baseline. 
However, this lack of significance does not establish equivalence among the prompting conditions.}

The prompting strategies nevertheless differ in observed cost.
\rev{Due to higher token usage, \textit{Chain-of-Thought} has the highest total and per-repair cost (CAD 0.000575), compared with the baseline (CAD 0.000236) and No-Context (CAD 0.000215), without significantly improving Repair@1. 
The ablation variants cost less than the baseline without significant Repair@1 differences, suggesting a more favorable observed cost-effectiveness trade-off for lightweight prompts.}

\begin{tcolorbox}[colback=gray!10, colframe=black, boxrule=0.5pt]
\textbf{Summary of Exploratory Extension.}
Overall, we observe that feedback effectiveness may vary by mutation category. 
For logical mutations, no pairwise difference remains significant after correction; meanwhile, for compiler mutations, \textit{compiler} and \textit{simple} feedback outperform simulated \textit{human} feedback but not each other. 
Finally, prompting variants show no significant Repair@1 differences, while lightweight prompts reduce cost relative to \textit{Chain-of-Thought}.
\end{tcolorbox}

\section{Discussion}
\label{sec-discussion}

In this section, we further discuss some insights and implications derived from our study replication and its exploratory extension for Java.

\subsection{Feedback Patterns and Prompting Trade-offs}
\label{sec:discussion-compiler}

Our exploratory Java results reveal mutation-dependent patterns that should be interpreted cautiously. 
Unlike the Python CoderEval results, where \textit{test} feedback performs best (Table~\ref{tab:rq2-results}), \textit{compiler} feedback achieves the highest observed Repair@1 for compiler mutations, although it does not differ significantly from \textit{simple} feedback.
The locations, symbols, and type information reported by \texttt{javac} may support localized repairs, while the strong performance of \textit{simple} feedback suggests that some mutations are repairable from code context alone. 
For logical mutations, \textit{simple} feedback ranks first, but does not differ significantly from \textit{test} feedback. 
One possible explanation is that JUnit~4.13.2 provides limited guidance when assertions lack customized messages, making the code and task description more informative than
the test output.

\rev{Prompting strategies differ more clearly in cost than in repair effectiveness. 
Increased prompt and response verbosity raises token usage~\cite{renze2024benefits,pan2025hidden}: \textit{Chain-of-Thought} is the least cost-efficient strategy, while \textit{Few-shot} adds input overhead; neither produces a statistically supported Repair@1 gain. 
Conversely, the ablation variants reduce cost while showing no statistically supported Repair@1 differences from the baseline, favoring lighter prompts in this setting.}

\subsection{Implications for Multilingual Repair Evaluation}
\label{sec:discussion-transfer}

Our Java exploratory extension should not be interpreted as a direct Java--Python comparison.
A rigorous comparison across programming languages requires controlling different dimensions, including the repair model, evaluation metric, source of bugs, mutation strategies, test quality, and feedback format.
In this study, the Java extension relies on LLM-generated mutations, whereas FeedbackEval combines tool-based mutations, LLM-generated errors, and incorrect code generations.
Such a mismatch may impact repair difficulty and influence the observed effectiveness of different feedback types.

In our study, the mutation strategy represents an important confound.
Our focus on LLM-generated mutations is motivated by recent studies showing that LLMs can generate more realistic and semantically meaningful faulty variants, complementing traditional mutation tools that often rely on predefined syntactic transformations~\cite{tip2025llmorpheus,wang2026comprehensive}.
However, since the mutations are generated by an LLM, they may reflect error patterns that are more familiar to the same or similar LLMs used during repair.
As a result, LLM-generated bugs may be easier for the repair model to recognize and fix.
Future multilingual repair benchmarks should therefore apply consistent mutation strategies across languages, use independent models or rule-based tools for mutation generation, and report mutation-difficulty indicators such as edit distance, changed control-flow structures, affected variables, and failure-triggering test characteristics.
However, these limitations do not invalidate the Java extension; they rather clarify its role: our extension provides exploratory evidence that feedback effectiveness can vary under a different benchmark, mutation strategy, and programming language ecosystem.
Furthermore, we also consider a smaller and less capable LLM.
Finally, these factors motivate and reveal the need for more controlled multilingual benchmarks before drawing strong conclusions about cross-language generalization in LLM-based repair.

\subsection{Replication Challenges of LLM-based Studies}
\label{sec:replication_challenges}

Replicating LLM-based studies presents several practical and methodological challenges that differ from traditional software engineering experiments \cite{wagner2025towards}. 
During our replication of FeedbackEval, we encountered different issues related to API variability, computational cost, missing experimental details, and evolving model availability.

\textbf{Resource Constraints and Computational Cost.}
Running the full FeedbackEval benchmark requires a large number of API calls. 
For RQ1 and RQ2 alone, the benchmark requires executing 394 tasks across four feedback modalities and two models, resulting in more than 3,000 API requests.
Additional prompting experiments further increase this number substantially. 
Each request incurs both financial cost and execution latency, making large-scale replications computationally expensive \cite{shekhar2024towards}. 
These costs create practical constraints for researchers attempting to replicate LLM-based experiments.

\textbf{Implications for Statistical Reporting.}
Our experience also highlights the need for stronger statistical reporting in LLM-based replication studies.
Because Repair@1 values can differ by only a few percentage points across feedback or prompting variants, descriptive percentages alone can lead to overinterpretation. 
In our extension, some apparent differences, such as simple versus \textit{test} feedback for Java logical mutations, are small and not statistically significant after correction.
Future replication studies should therefore report confidence intervals, paired or aggregate statistical tests when appropriate, and corrections for multiple comparisons.

\textbf{Model Deprecation.} Another practical challenge concerns the rapid evolution of LLMs \cite{baltes2025guidelines}. 
During this study, the model \texttt{claude-3-5-sonnet-20241022} was officially deprecated.\footnote{https://platform.claude.com/docs/en/about-claude/model-deprecations} 
All replication experiments involving this model were therefore executed before its deprecation date to maintain consistency with the original study. 
This highlights an additional difficulty in reproducing LLM-based studies: the underlying models may change or disappear over time, directly limiting long-term reproducibility \cite{ma2024schrodinger}.

\section{Related Work}
\label{sec-related_work}

Recent advances in LLM-based program repair have explored the role of feedback mechanisms, cross-language transfer, and structured representations for improving repair effectiveness. 
In this section, we position our work regarding prior research, highlighting how our study extends existing findings by focusing on the reproducibility of feedback-driven repair and on how feedback strategies behave in an exploratory Java repair setting.

Prior work has also highlighted the challenges of generalization in APR. 
Zirak and Hemmati \cite{zirak2024improving} investigate the impact of domain shift in repair tasks, where models trained and evaluated on similar projects exhibit significant performance degradation when applied to new domains.
To address this issue, the authors propose a new framework that adapts to the target project while improving repair effectiveness.
Similar to our goal, this work highlights that repair performance is highly sensitive to differences in data distribution and problem characteristics. 
In contrast, our work investigates the reproducibility of feedback-driven repair results and explores how feedback strategies behave in a Java repair setting.

In the context of LLM-based repair, prior work has also explored the ability of LLMs to transfer code repair capabilities across programming languages.
Baltaci et al.~\cite{baltaji2025cross} show that transfer learning across multiple source and target languages significantly outperforms zero-shot approaches for code repair tasks. 
Similarly, LANTERN~\cite{luo2025unlocking} exploits differences in language-specific model proficiency by translating code into languages where LLMs perform better, improving repair performance.
While these studies focus on improving repair effectiveness through cross-language transfer or translation, they do not investigate how different feedback modalities behave in a replication and extension setting. 
In contrast, we partially replicate a feedback-driven repair benchmark and conduct an exploratory Java extension to examine how feedback modalities behave under a different programming-language ecosystem.

Several recent approaches have investigated how structured feedback can improve LLM-based program repair.
DSrepair~\cite{ouyang2025knowledge} leverages AST-based feedback to localize errors more precisely, achieving improvements in repair performance while reducing token usage. 
Similarly, RePair \cite{zhao2024repair} introduces reward models that act as virtual compilers to guide iterative repair through feedback signals.
These approaches demonstrate the potential of structured and iterative feedback mechanisms to improve repair effectiveness.
However, they primarily focus on specific domains (e.g., data science code) or iterative repair settings. 
In contrast, our study evaluates multiple feedback strategies in a controlled, single-iteration setting and examines their behavior in a Java repair extension.
This complements prior work by highlighting how feedback representation, tooling, and mutation strategy may shape observed repair outcomes.

Finally, in the context of vulnerability repair tasks \cite{zhou2025large}, recent work has explored both feedback integration and feedback representation.
In this context, Kulsum et al. \cite{kulsum2024case} present \textit{VRpilot}, an LLM-based approach combining \textit{Chain-of-Thought} reasoning with iterative patch validation feedback from external tools like compilers, test suites, and sanitizers for C and Java programs. 
Their results show that integrating reasoning and validation feedback leads to improved repair performance.
Meanwhile, Zhang et al. \cite{zhang2024vuladvisor} propose \textit{VulAdvisor}, a framework responsible for generating natural-language repair suggestions to guide vulnerability fixing, rather than directly producing patches.
Their results show that providing more informative and human-readable feedback can improve repair effectiveness.
While these studies emphasize the importance of both feedback integration and feedback representation in LLM-based repair, they mainly focus on improving repair performance in specific scenarios and do not systematically compare multiple feedback types in a replication setting or examine how feedback behavior changes under a different repair benchmark and tooling ecosystem.

\section{Threats to Validity}
\label{sec-threats}

\textbf{Internal Validity.}
Java mutations were generated exclusively using GPT-4o-mini, whereas the original FeedbackEval Python dataset combines rule-based mutation operators, LLM-generated errors, and incorrect LLM-generated solutions. 
This difference may introduce a confound, as LLM-generated Java mutations may reflect error patterns closer to the model's training distribution, potentially making them easier for the repair model to fix. 
In contrast, rule-based mutations used in the Python dataset were designed to systematically stress repair capabilities. 
Consequently, the Java repair rates observed in our experiments may partially reflect mutation difficulty and benchmark-construction choices rather than language-intrinsic effects.
The Java experiments rely on JUnit~4.13.2 for test execution, which typically produces concise assertion failure messages. 
In contrast, the Python benchmark relies on \texttt{pytest}, whose failure output often includes richer diagnostic information. 
This difference may reduce the usefulness of \textit{test} feedback in Java and partially explain why structured \textit{test} feedback does not provide a clear advantage over simpler feedback in our Java setting.

\textbf{Construct Validity.}
Regarding the translation of test cases, the use of \textit{GPT-4o-mini} might alter test semantics or introduce inconsistencies, potentially biasing our results.
To mitigate this threat, we manually inspected all 50 translated JUnit tests and found no semantic inconsistencies with the original assertion scripts. 
We also validated that the reference implementations pass the translated tests and that the injected mutations fail them.
Repair success is defined as code that compiles and passes the associated test suite.
This binary definition does not capture other important aspects of software quality, such as readability, maintainability, or the introduction of subtle semantic bugs not covered by the tests.
Human feedback is simulated using an LLM rather than being collected from human developers.
Although this allows controlled experiments, simulated feedback may not reflect the priorities, expertise, or contextual knowledge of real developers.

Unlike the original study, our extension relies exclusively on LLM-generated mutations.
While this design choice enables controlled and consistent mutation generation across tasks, it may introduce bias toward errors that are more representative of the LLM's training distribution.
As a result, these mutations may be easier for the same or similar models to repair, potentially inflating Repair@1 scores.
To partially mitigate this limitation, our dataset includes a diverse distribution of 13 logical mutation types, covering a range of common error patterns, such as incorrect conditions, off-by-one errors, and wrong return values.
Nevertheless, the mutations remain LLM-generated and may not fully reflect the characteristics of real-world or tool-generated bugs.

\textbf{External Validity.}
Our replication evaluates a subset of the models considered in the original study.
While the evaluated models are strong proprietary LLMs, this choice may limit the generalizability of our findings to other LLMs with different capabilities, architectures, or training data.
As prior work suggests that LLM behavior can vary across models, the observed feedback effectiveness patterns may not fully transfer to all model families.
Regarding our extension, our evaluation uses a single model (GPT-4o-mini) on 50 unique CoderEval tasks, resulting in 100 mutated repair instances.
Overall, we prioritize controlled experimental conditions over scale, enabling focused analysis of feedback modalities across mutation categories.

Although this setup enables controlled analysis within the Java setting, the results may not generalize to other LLMs, larger datasets, or real-world bugs involving multiple files and complex software architectures.
Additionally, our single-iteration repair setup does not capture iterative debugging workflows commonly used in practical repair systems.
\rev{Mutation generation, JUnit test construction, simulated human-feedback generation, and repair generation were all performed using the same LLM.}
This may introduce bias, as the model could generate mutants and tests that reflect patterns within its own training distribution, making them easier to recognize and repair.
Future work could mitigate this threat by generating mutants and tests using independent models or rule-based mutation frameworks.

\textbf{Conclusion Validity.}
For the Java extension, the same tasks are evaluated under each feedback and prompting condition.
We therefore analyze paired task-level outcomes using Cochran's \(Q\) and two-sided exact McNemar tests with Holm correction, alongside Wilson 95\% confidence intervals.
Each condition was executed once; therefore, we do not quantify stochastic variability across repeated generations.
Comparisons between our Python replication and the published FeedbackEval results remain descriptive.
Small or non-significant differences should be interpreted cautiously.

\section{Conclusion}
\label{sec-conclusion}

In this study, we partially replicated and extended the FeedbackEval benchmark \cite{dai2025feedbackeval} to examine the reproducibility of feedback-driven LLM-based code repair, while exploring how different feedback types behave in a Java repair setting. 
Regarding our replication of the original Python-based study, our results largely confirm the reported trends and findings, particularly the effectiveness of test-based feedback over alternative types, supporting the robustness of the original findings under our experimental setup.

Building on this, we moved forward with our extension to Java, revealing interesting findings. 
\rev{In our Java extension, \textit{simple} feedback ranks highest for logical mutations, although no pairwise difference remains significant after correction. 
For compiler mutations, \textit{compiler} and \textit{simple} feedback outperform simulated \textit{human} feedback but do not differ significantly from each other.}
Such a divergence appears to be influenced by differences in feedback types and tooling characteristics, such as the verbosity of test frameworks.
We must emphasize that our findings should be interpreted considering our experimental design. 
Differences in mutation generation and dataset construction between Python and Java may affect repair difficulty and consequently influence comparisons with prior Python findings. 
Therefore, our results highlight observed behavioral differences under controlled conditions rather than isolating purely language-intrinsic effects.
\rev{Finally, lightweight prompts show a more favorable observed effectiveness--cost trade-off in our setting.}

Overall, our study reinforces the value of replication in LLM-based software engineering research, while highlighting the need for more controlled multilingual repair benchmarks.
Future work should investigate consistent mutation-generation strategies across languages and different mutation sources, and explore richer feedback representations.
Finally, extending this study to additional programming languages, models, and real-world repair scenarios would further improve our understanding of feedback-driven LLM-based program repair.

\section*{Replication Package and Artifacts}
We provide two main artifacts.
First, the \textbf{replication package}, including scripts, prompts, and datasets for reproducing the original FeedbackEval results, is available online \cite{replication_appendix}.
Second, the \textbf{extension repository} provides the Java tasks, feedback artifacts, and instructions for obtaining the Docker evaluation image~\cite{extension_appendix}.

\section*{AI-Assisted Content Disclosure}

LLMs were used to assist the authors with grammar revision and to support some scripts used in this study. All outputs were reviewed and validated by the authors, who are fully responsible for the content presented in this study.

\section*{Acknowledgments}
This work was partially supported by the Natural Sciences and Engineering Research Council of Canada, the Canadian Institute for Advanced Research, and the Canada Research Chairs Program.

\bibliographystyle{IEEEtran}
\bibliography{sample-base}

\end{document}